\documentclass{ws-procs9x6}
\usepackage{epsfig,graphicx}
\usepackage{amsmath,amsfonts,amssymb}
\usepackage{citesort}

\def\neumass{m_{\tilde\chi_1^0}}
\newcommand{\crosssec}{\sigma_{\tilde\chi^0_1-p}}

\def\neut{\tilde\chi_1^0}

\def\lsim{\raise0.3ex\hbox{$\;<$\kern-0.75em\raise-1.1ex\hbox{$\sim\;$}}}
\def\gsim{\raise0.3ex\hbox{$\;>$\kern-0.75em\raise-1.1ex\hbox{$\sim\;$}}}
\def\nmh{{\sc nmhdecay}}

\begin{document}

\title{Direct detection of neutralino dark matter in the NMSSM}

\author{Ana M. Teixeira}

\address{Departamento de F\'{\i}sica Te\'{o}rica C-XI and Instituto
de F\'{\i}sica Te\'{o}rica C-XVI, \\
Universidad Aut\'{o}noma de Madrid,
Cantoblanco, E-28049 Madrid, Spain\\
E-mail: teixeira@delta.ft.uam.es
}

\maketitle

\abstracts{
We address the direct detection of neutralino dark matter in the framework of
the Next-to-Minimal Supersymmetric Standard Model.
We conduct a detailed
analysis of the parameter space, taking into account all the available
constraints from LEPII, and compute the
neutralino-nucleon cross section. We find that sizable values for the
detection cross section, within the reach of dark matter detectors,
are attainable in this framework, and are associated with 
the exchange of very light Higgses,
$m_{h_1^0}\lesssim 70$ GeV, the latter exhibiting a significant
singlet composition. 
} 

\section{Introduction}\label{intro}
Supersymmetric (SUSY) theories offer some excellent candidates for 
dark matter. In particular, the lightest neutralino, $\neut$, is the
leading one within the class of Weakly Interacting Massive Particles
(WIMPs). WIMPs can be directly detected via elastic scattering on
target nuclei and there are currently a large number of experiments
devoted to the direct detection of WIMP dark matter\cite{review}. 

We have studied the theoretical predictions for the
direct detection of neutralino dark matter in the framework of the
Next-to-Minimal Supersymmetric Standard Model (NMSSM)\cite{Cerdeno:2004xw}. 
Via the
introduction of a singlet superfield $S$, the NMSSM offers an elegant
solution to the $\mu$ problem of the Minimal Supersymmetric Standard
Model (MSSM), while at the same time it
also renders the Higgs ``little fine tuning problem'' of the MSSM less
severe. 
The new fields in the model
mix with the corresponding MSSM ones, giving rise to a richer and more complex
phenomenology. In particular, a very light neutralino may be present
and a very light Higgs boson is not experimentally 
excluded\cite{Ellwanger:2003jt}. The
latter aspects, among other features, 
may modify the results concerning the neutralino-nucleon cross
section with respect to those of the MSSM.

\section{Neutralino-nucleon cross section in the NMSSM}
The NMSSM is defined by the following superpotential
\begin{equation}
W=
Y_u \, H_2\, Q\, u +
Y_d \, H_1\, Q\, d +
Y_e \, H_1\, L\, e 
- 
\lambda \,S \,H_1 H_2 +\frac{1}{3} \kappa \,S^3\,,
\end{equation}
with $S$ a singlet under the
Standard Model (SM) gauge group.
After spontaneous electroweak symmetry breaking, the neutral Higgs scalars
develop vacuum expectation values,
$\langle H_1^0 \rangle = v_1, \, 
\langle H_2^0 \rangle = v_2, \, 
\langle S \rangle = s$,
and an effective $\mu$ term is thus generated, $\mu \equiv
\lambda s$. 

In the absence of CP violation in the Higgs sector, the CP-even and
CP-odd states do not mix. In the NMSSM, we find three scalar and
two pseudoscalar  Higgs states. Of particular relevance to our analysis is
the lightest scalar, $h_1^0$, which can be written in terms of
the original fields as
\begin{equation}
h_1^0= S_{11} H_1^0 + S_{12} H_2^0+  S_{13} S\,,
\end{equation}
where $S_{ab}$ diagonalises the $3 \times 3$ scalar Higgs mass
matrix\cite{Cerdeno:2004xw}. 
In the neutralino sector, 
the singlino ($\tilde S$) mixes with the Bino, Wino and Higgsinos, and in
this model, the lightest neutralino can be
expressed as the combination
\begin{equation}
\tilde \chi^0_1 = N_{11} \tilde B^0 + N_{12} \tilde W_3^0 +
N_{13} \tilde H_1^0 + N_{14} \tilde H_2^0 + N_{15} \tilde S\,,
\end{equation}
with $N$ the matrix that diagonalises the $5 \times 5$ neutralino mass 
matrix\cite{Cerdeno:2004xw}.

The leading contributions to the neutralino-nucleon cross section
($\crosssec$) are
associated with the scalar, spin- and velocity-independent term 
$\alpha_3 \bar{\tilde \chi}^0_1 {\tilde \chi}^0_1\bar q\,q$ in the
effective Lagrangian\cite{review},
which receives contributions from squark and Higgs
exchange diagrams\cite{Cerdeno:2004xw,nmssm:old}. 
The term $\alpha_{3i}^{\tilde q}$ is formally identical to the MSSM case,
differing only in the new neutralino mixings stemming from the
presence of a fifth component, and plays a sub-leading role in our
analysis. Regarding the Higgs mediated interaction term
($\alpha_{3i}^{h}$), the situation is
slightly more involved since both vertices and the exchanged Higgs
scalar 
significantly reflect the new features of the NMSSM\cite{Cerdeno:2004xw}.
It should be emphasised that the exchange of light Higgs scalars in
the $t$-channel might provide a considerable enhancement to the
neutralino-nucleon cross section.

\section{Results and discussion}
In our study\cite{Cerdeno:2004xw}, 
we were particularly interested in the various
NMSSM scenarios which might potentially lead to values
of $\crosssec$ in the sensitivity range of detectors which are 
currently running or in preparation.
The analysis of the NMSSM parameter space (minimization of the
potential, computation of spectrum and compatibility with LEP experimental 
constraints) was done using the program \nmh~\cite{Ellwanger:2004xm}. 

At the electroweak scale, we have the following set of free, independent
parameters: 
$\lambda$, $\kappa$, $\mu(=\lambda s)$, $\tan \beta$, the
soft trilinear terms for the Higgs scalars, $A_\lambda$, $A_\kappa$,
the soft gaugino masses $M_i$, and a common SUSY scale for the
remaining squark masses and trilinear couplings, $M_{\text{SUSY}}$.

Not only the cross section itself, but also the allowed regions of the
low-energy NMSSM parameter space are strongly sensitive to variations
of the input parameters. 
It proved very illustrative to analyse the relevant features of the
model in the plane generated by the Higgs couplings in the
superpotential, $\lambda$ and $\kappa$.

\begin{figure}[t]
  \epsfig{file=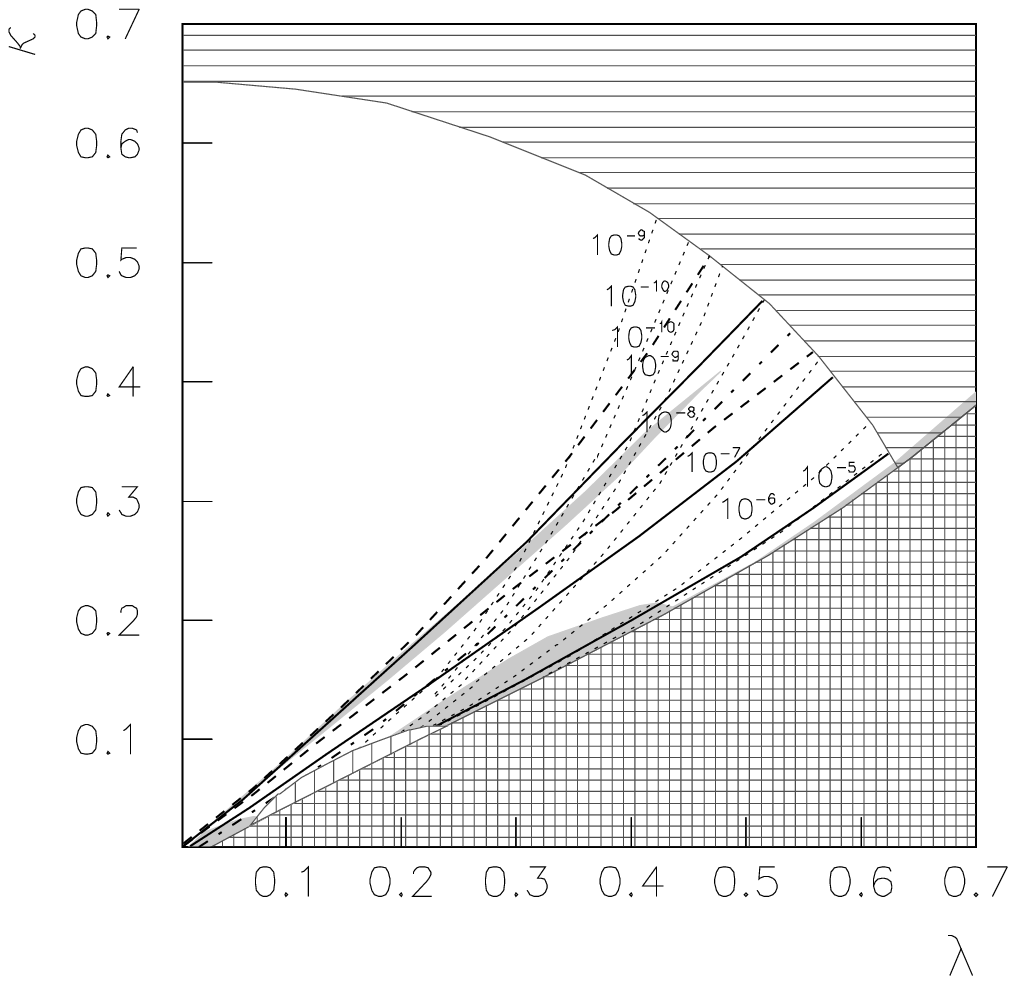,height=52mm}
  \hspace*{-5mm}\epsfig{file=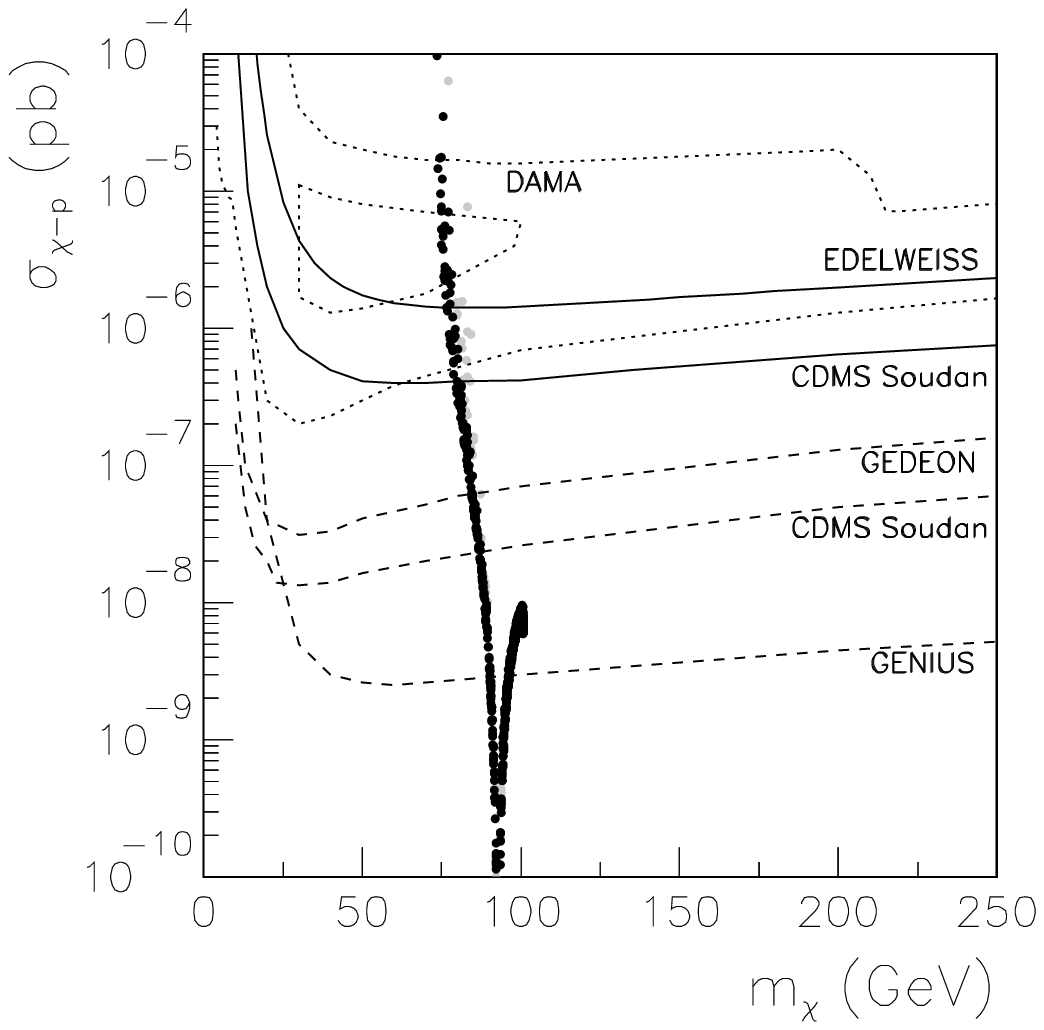,height=52mm}
  \caption{On the left,
    $(\lambda,\kappa)$ parameter space with the corresponding
    constraints for the case
    $A_\lambda=200$ GeV, $A_\kappa=-200$ GeV,  $\mu=110$ GeV  and
    $\tan\beta=3$.
    Shaded areas are excluded, while  
    dotted lines in the accepted region
    represent contours of $\crosssec$.
    From top to bottom, solid lines indicate different
    values of lightest scalar Higgs mass,
    $m_{h_1^0}=114,\,75,\,25$ GeV, dashed lines separate the regions where the
    lightest scalar Higgs has a singlet composition given by
    ${S_{13}^{\,2}}=0.1,\,0.9$ and a dot-dashed line reflects the singlino
    composition of the lightest neutralino (below/above $N_{15}^2=0.1$).
    On the right, and for the same choice of input parameters, 
    scatter plot of $\crosssec$ 
    as a function of the neutralino
    mass, with the sensitivities of present and projected experiments.
    Black dots fulfil all constraints, while in grey are those
    experimentally excluded.
    }
  \label{++-ka}
\end{figure}

As an example, we plot in Fig.\ref{++-ka}
the $(\lambda,\kappa)$ parameter space and the cross section versus
the lightest neutralino mass for $\tan\beta=3$,
$A_\lambda=200$ GeV, $A_\kappa=-200$ GeV and  $\mu=110$ GeV (taking 
$M_2=2 M_1= M_{\text{SUSY}}=1$ TeV).
Significant regions of the parameter space are
excluded due to theoretical and experimental constraints. The first
class comprises the presence of tachyonic CP-even Higgs scalars
(gridded area) and the occurrence of false minima and Landau poles (vertically
and horizontally ruled regions, respectively). 
The grey area is associated to points that do not satisfy the LEP constraints.
As can be clearly seen from both plots, very large values of 
$\crosssec$ (in fact, even points already excluded by direct searches)
can be obtained. From the inspection of the $(\lambda,\kappa)$ plane,
it is clear that such large values are associated with very light
Higgs states (as light as 20 GeV), which are experimentally viable due
to their important singlet character ($0.9\lsim
S_{13}^{\,2}\lsim0.95$). 
In this case, the NMSSM nature is clearly patent in
the compositions of the lightest neutralino (a mixed singlino-Higgsino
state) and of $h_1^0$ (light, and mostly singlet-like).

\begin{figure}
  \epsfig{file=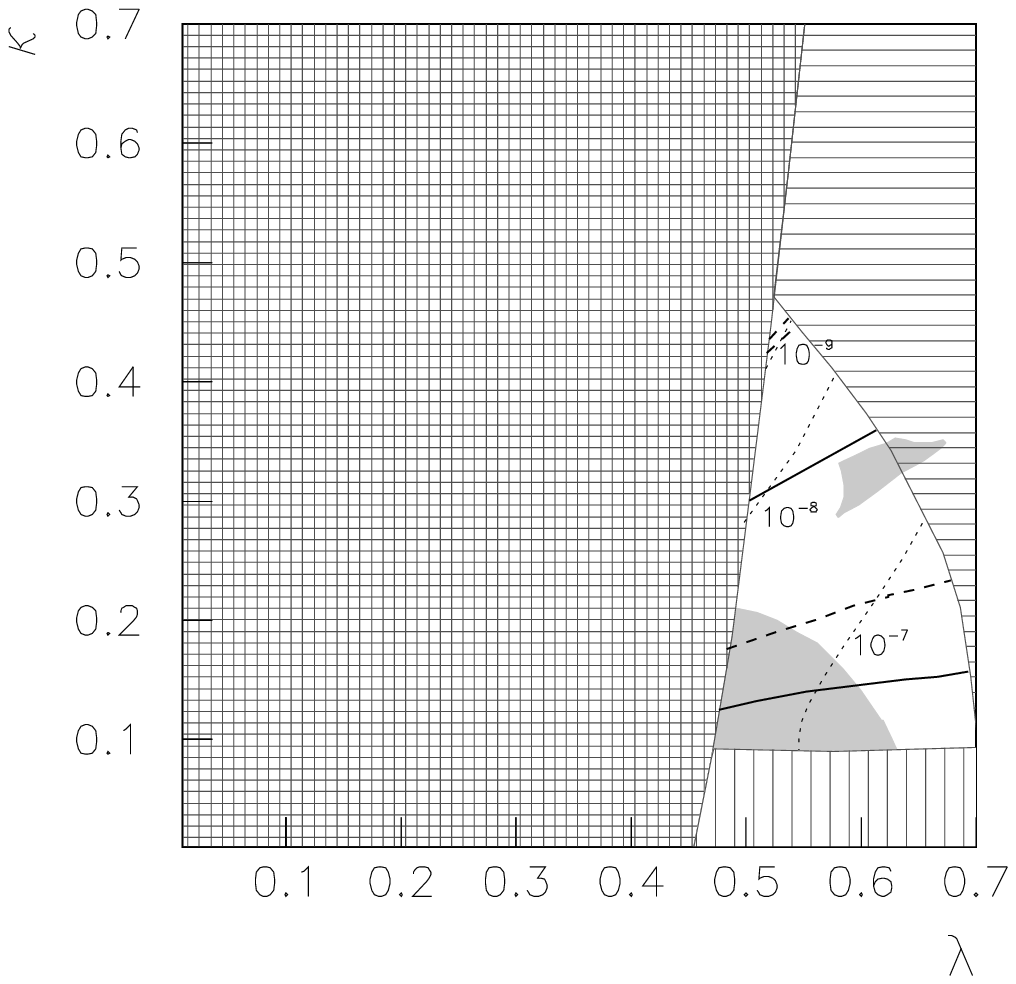,height=52mm}
  \hspace*{-5mm}\epsfig{file=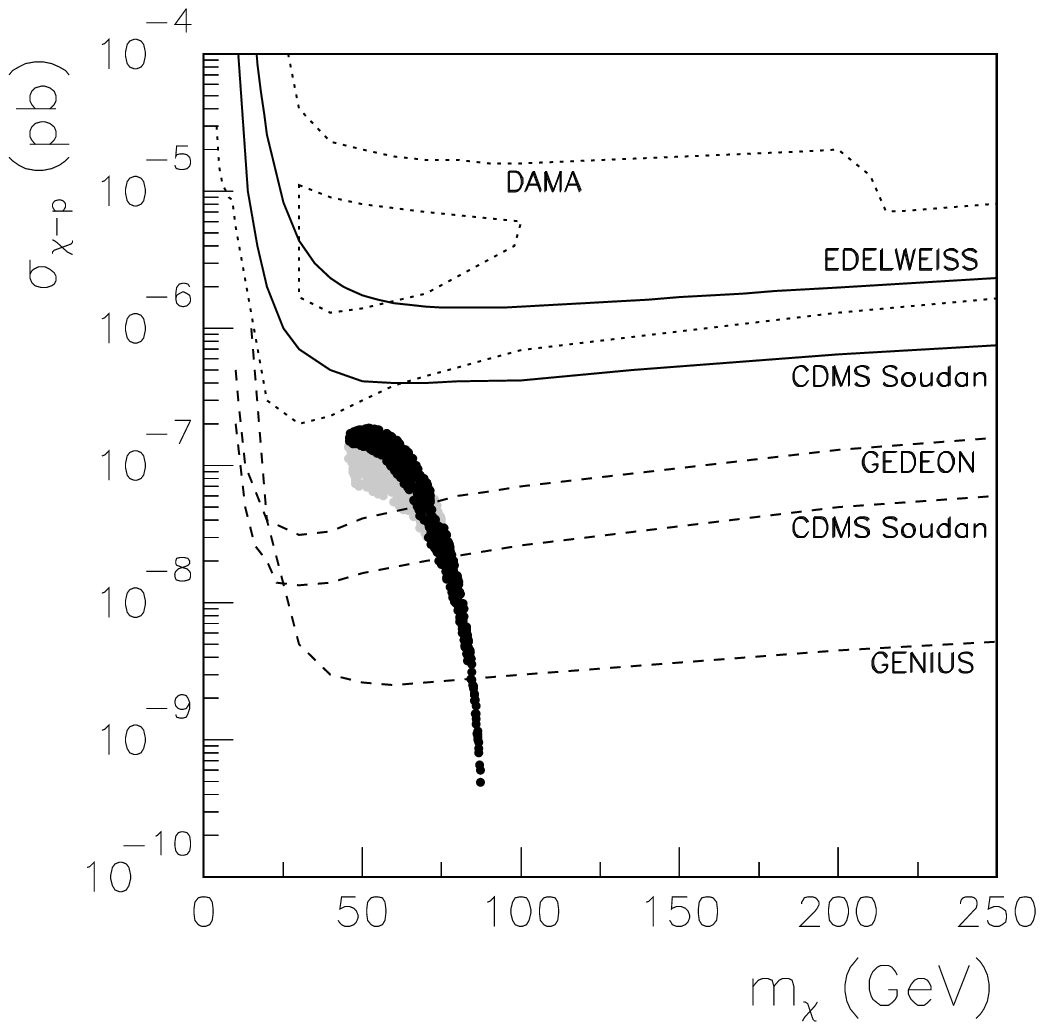,height=52mm}
  \caption{The same as in Fig.\ref{++-ka} but for
    $A_\lambda=300$ GeV, $\mu=110$
    GeV, $A_\kappa=50$ GeV,
    and $\tan\beta=3$. Only the lines with $m_{h_1^0}=114,\,75$~GeV are
    represented. Regarding the neutralino composition, only the line
    with $N_{15}^2=0.1$ is shown.
} 
  \label{+++al}
\end{figure}

Another example, but for a distinct region in the NMSSM parameter
space, is depicted in Fig.\ref{+++al}, for $A_\lambda=300$ GeV,
$\mu=110$~GeV, $A_\kappa=50$ GeV, and $\tan\beta=3$.
In this case, tachyons arise in both CP-even and CP-odd sectors, and
close to these areas the experimental constraints from the Higgs
sector are more severe. Nevertheless, regions
with very light Higgses and $\neut$ are experimentally viable.
In particular, neutralinos with an important singlino composition,
$N_{15}^2\lsim0.45$, can be obtained with $\neumass\gsim45$ GeV.
Moreover, the lightest Higgses ($m_{h^0_1}\approx65-90$ GeV) are all
singlet-like, and this favours large values of the cross section
($\crosssec\lsim2\times10^{-7}$ pb).

Until here we have only addressed cases where $\neut$ is essentially a
singlino-Higgsino mixture, and this stems from the chosen hierarchy, 
$\mu\,<\,M_1\,<M_2$. 
By relaxing the latter, more general compositions for the neutralino
can be found. 
Let us go back to the example already analysed in Fig.\ref{++-ka},
but now taking two different values for $\mu$,
$\mu=200,\,500$ GeV. Regarding the gaugino
masses, we allow variations in the Bino mass as $50$ GeV $\leq
M_1\leq 500$ GeV, with the GUT relation
$M_1=\frac{1}{2}\,M_2$. 
 \begin{figure}[!t]
\epsfig{file=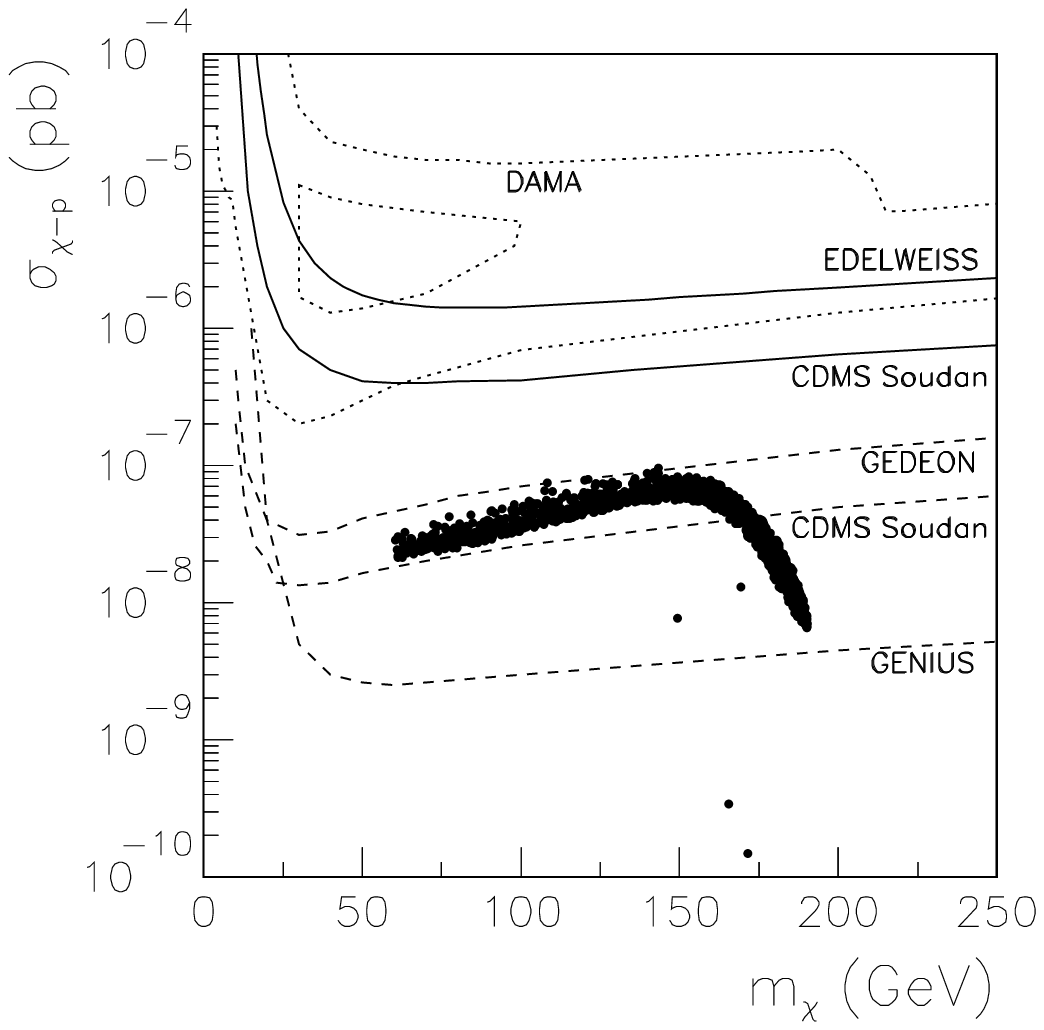,height=52mm}  
  \hspace*{-5mm}\epsfig{file=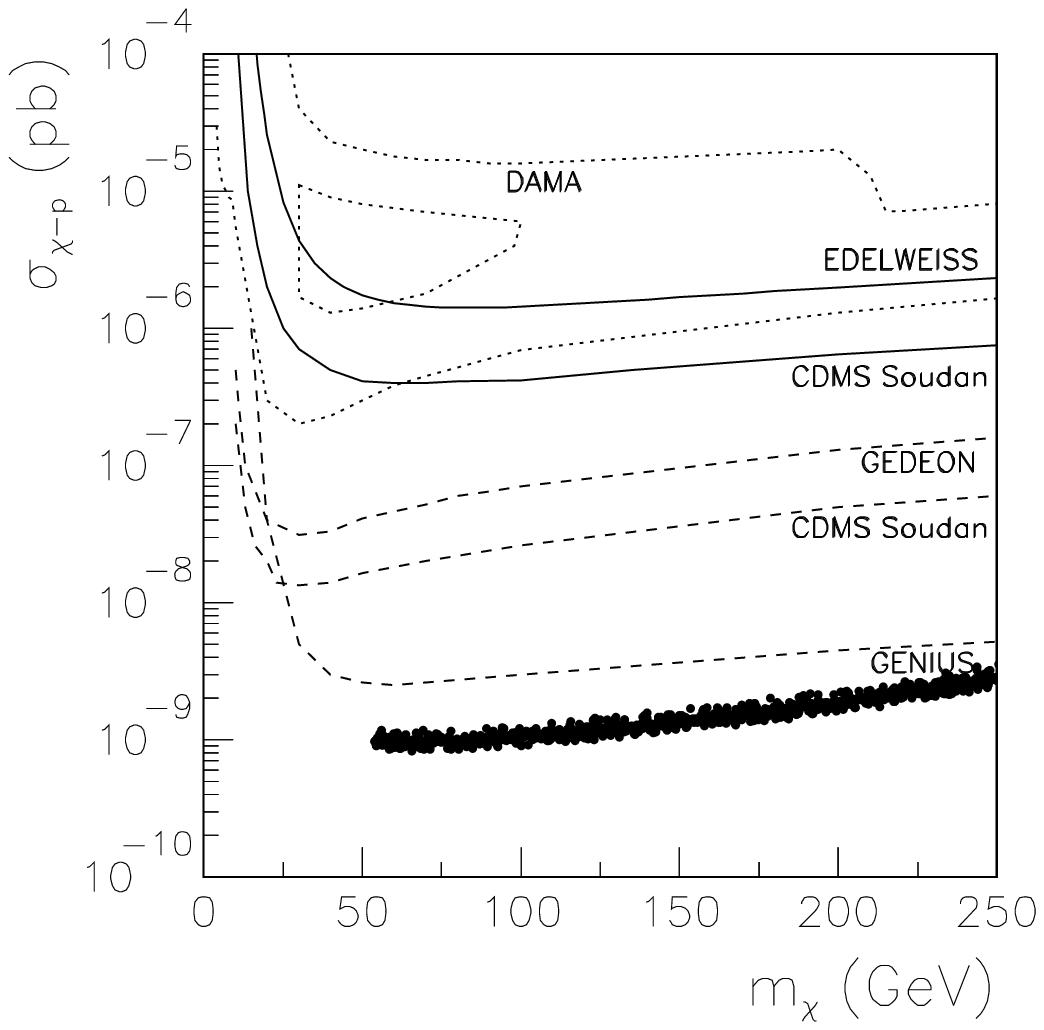,height=52mm}
  \caption{Scatter plot of the scalar neutralino-nucleon cross
    section as a function of $\neumass$ for $A_\lambda=200$
    GeV, $A_\kappa=-200$ GeV, $\tan\beta=3$, $\mu=200,\,500$
    GeV. The gaugino masses satisfy the GUT
    relation, with $M_1$ in the range $50$ GeV $\leq M_1\leq 500$ GeV. }
  \label{gutxx}
\end{figure}
As the value of $\mu$
increases, so does the gaugino composition of $\neut$.
Simultaneously, the lightest Higgs becomes heavier and more
doublet-like. For $\mu=500$ GeV, neutralinos lighter than
$\neumass\lsim 375$ GeV are all Bino-like. 
The Higgs mediated interaction is now negligible and
detection only takes place through the squark mediated
interaction. In contrast to the previous examples, 
Bino-like neutralinos would have
$\crosssec\lsim10^{-9}$ pb, and thus would be beyond the sensitivities
of even the largest projected dark matter detectors.

Finally, it is important to conduct a more general survey of the NMSSM
parameter space in order
to obtain a global view on the theoretical
predictions for $\crosssec$, and their compatibility with present and
projected detectors.
We focus on the case $\mu=110$ GeV, with heavy gaugino
masses, $M_1=\frac12\,M_2=500$ GeV, since this choice leads to larger
predictions for $\crosssec$.
The rest of the input
parameters are allowed to vary in the ranges
$-600$~GeV~$\leq~A_\lambda~\leq~600$~GeV,
$-400$~GeV~$\leq~A_\kappa~\leq~400$~GeV, and we take 
$\tan\beta=2,\,3,\,4,\,5,\,10$, with
$\lambda,\,\kappa \in[0.01,0.8]$. 
\begin{figure}
  \epsfig{file=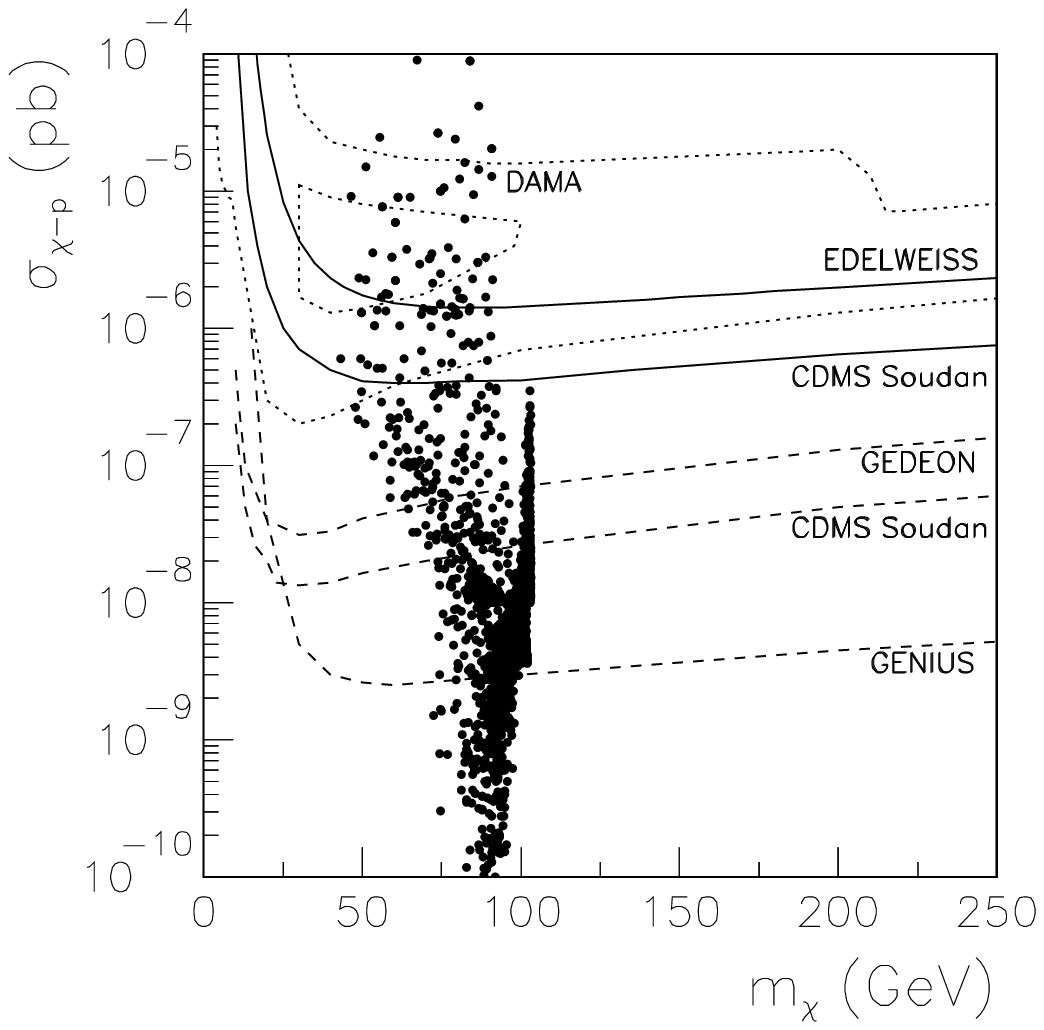,height=48mm}
  \hspace*{-3mm}\epsfig{file=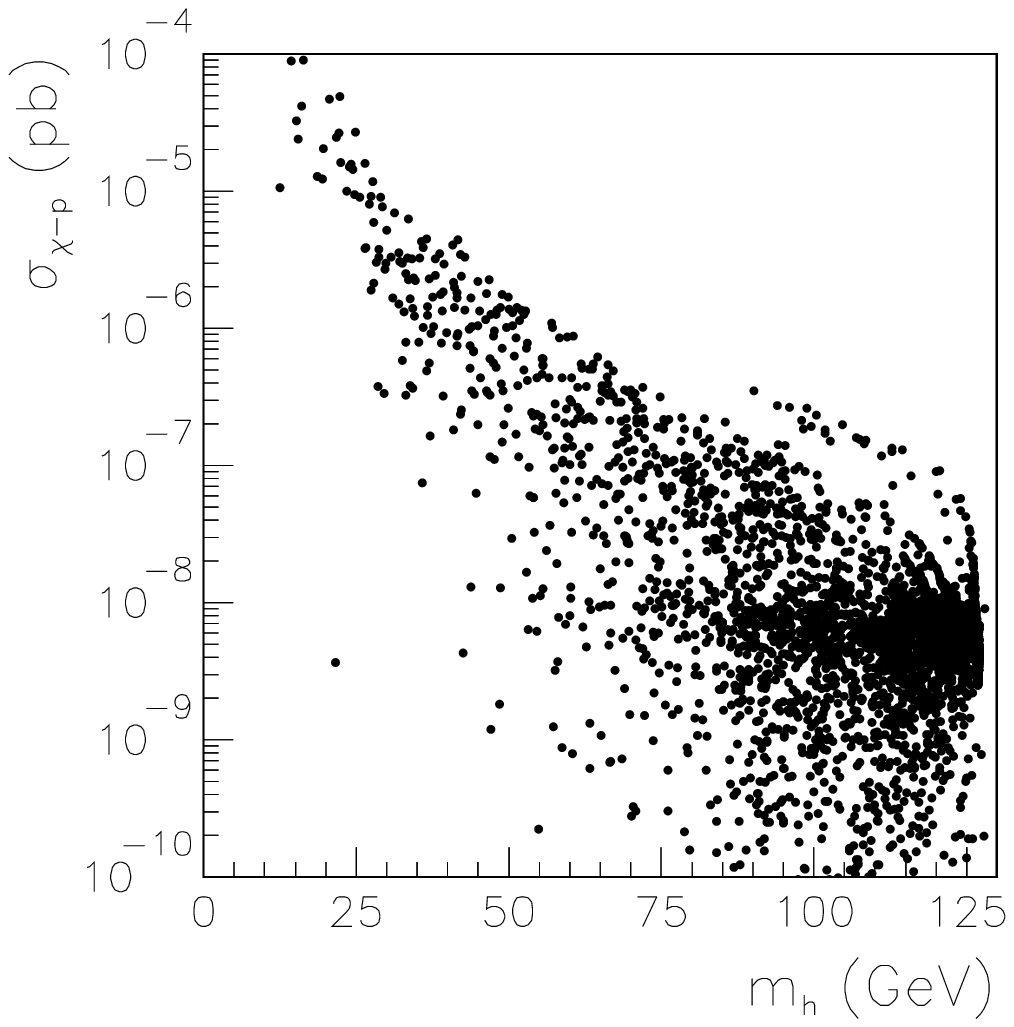,,height=48mm}
  \caption{Scalar neutralino-nucleon cross
    section as a function of the lightest neutralino mass (left) and 
    of the lightest Higgs mass (right). The input parameters are 
    $\mu=110$ GeV, 
    $-600$ GeV $\leq A_\lambda\leq600$ GeV,
    $-400$ GeV $\leq A_\kappa\leq400$ GeV, and $\tan\beta~=~2,\,3,\,4,
    \,5,\,10$.}
  \label{summarya}
\end{figure}
The results of this scan are shown in Fig.\ref{summarya}. 
Points with
large predictions for $\crosssec$ are found, and correspond to very light
singlet-like Higgses, even with $m_{h_1^0}\gsim15$ GeV, 
which are more easily
obtained for low values of $\tan\beta$ ($\tan\beta\lsim5$), and
typically originate from regions where $\mu A_\lambda>0$.

\section{Conclusions}

We have performed a systematic analysis of the low-energy parameter 
space of the
Next-to-Minimal Supersymmetric Standard Model (NMSSM),
studying the implications for the direct detection of neutralino dark
matter\cite{Cerdeno:2004xw}. 
In the computation of $\crosssec$ we have taken into account 
the relevant constraints on
the parameter space from accelerator data.
We have found that large values of $\crosssec$, even within the reach
of present dark matter detectors (see e.g. Fig.\ref{summarya}), 
can be obtained, 
and this is essentially due to the 
exchange of very light Higgses, $m_{h_1^0}\lsim 70$ GeV. 
The NMSSM nature is evidenced in this result, since such Higgses have 
a significant singlet composition, thus escaping detection and being in
agreement with accelerator data. 

\section*{Acknowledgments}
A.M. Teixeira acknowledges the support by
Funda\c c\~ao para a Ci\^encia e Tecnologia under
the grant SFRH/BPD/11509/2002.

\end{document}